\documentclass[article,twocolumn,amsmath,floats,showpacs,nofootinbib]{revtex4}
\usepackage{graphicx}
\usepackage{textcomp}
\usepackage{color,hyperref}
\hypersetup{colorlinks=true}
\begin{document}
\input epsf

\title{Point-contact spectroscopy of superconductors in the nonequilibrium state}

\author{I. K. Yanson, L. F. Rybal'chenko, N. L. Bobrov, and V. V. Fisun}
\affiliation{B.I.~Verkin Institute for Low Temperature Physics and
Engineering, of the National Academy of Sciences
of Ukraine, prospekt Lenina, 47, Kharkov 61103, Ukraine
Email address: bobrov@ilt.kharkov.ua}

\published {(\href{http://fntr.ilt.kharkov.ua/fnt/pdf/12/12-5/f12-0552r.pdf}{Fiz. Nizk. Temp.}, \textbf{12}, 552 (1986)); (Sov. J. Low Temp. Phys., \textbf{12}, 313 (1986)}
\date{\today}

\begin{abstract}A phase transition of the region of the superconductor near the point contact into a new nonequilibrium state at the critical density of nonequilibriumquasiparticles is observed.

\end{abstract}

\maketitle

In pure metallic point contacts, whose dimensions \emph{d} are smaller than the characteristic microscopic lengths of the material (electron and phonon mean-free paths, coherence lengths $\xi $ of the superconductor, etc.), electron and phonon states very far from thermal equilibrium are realized in the current-carrying state. If one or both electrodes are superconductors, then a nonequilibrium superconducting state arises near the microconstriction. Because of the rapid spreading of the current, the nonlinearities of the current-voltage characteristic (IVC) of the point contact give information about a very small region with dimensions of the order of $d\ll \xi $ near the contact, unlike tunneling contacts, whose dimensions are much larger than  . For this reason, it is possible to obtain additional information about nonequilibrium states in superconductors with the help of point contacts.\\
\begin{figure}[]
\includegraphics[width=0.47\textwidth,angle=0]{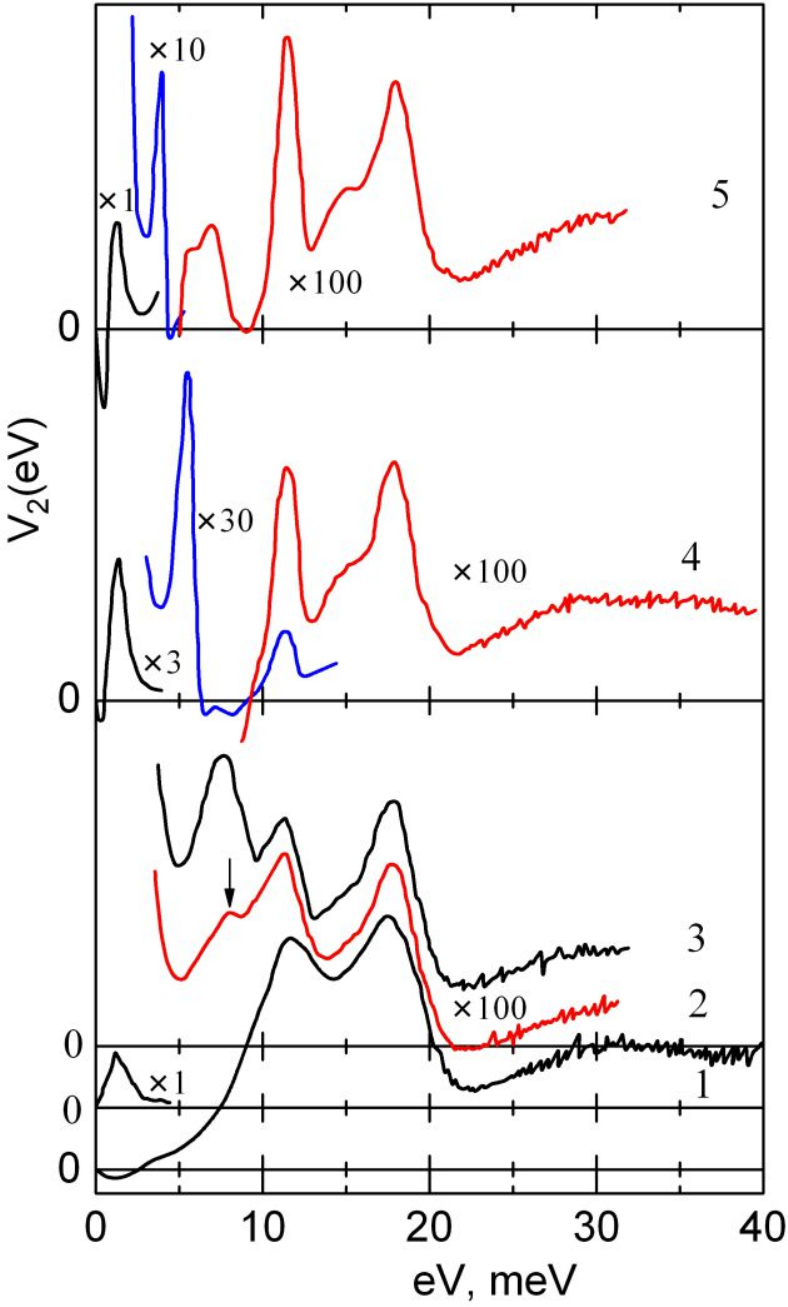}
\caption[]{Temperature evolution of the point contact spectra of $Ta$ ($Ta-Cu$ contact) at T, K: 4.8 (1); 3.9(2); 3.0(3); 2.25(4); 1,4(5).}
\label{Fig1}
\end{figure}
\begin{figure}[]
\includegraphics[width=0.47\textwidth,angle=0]{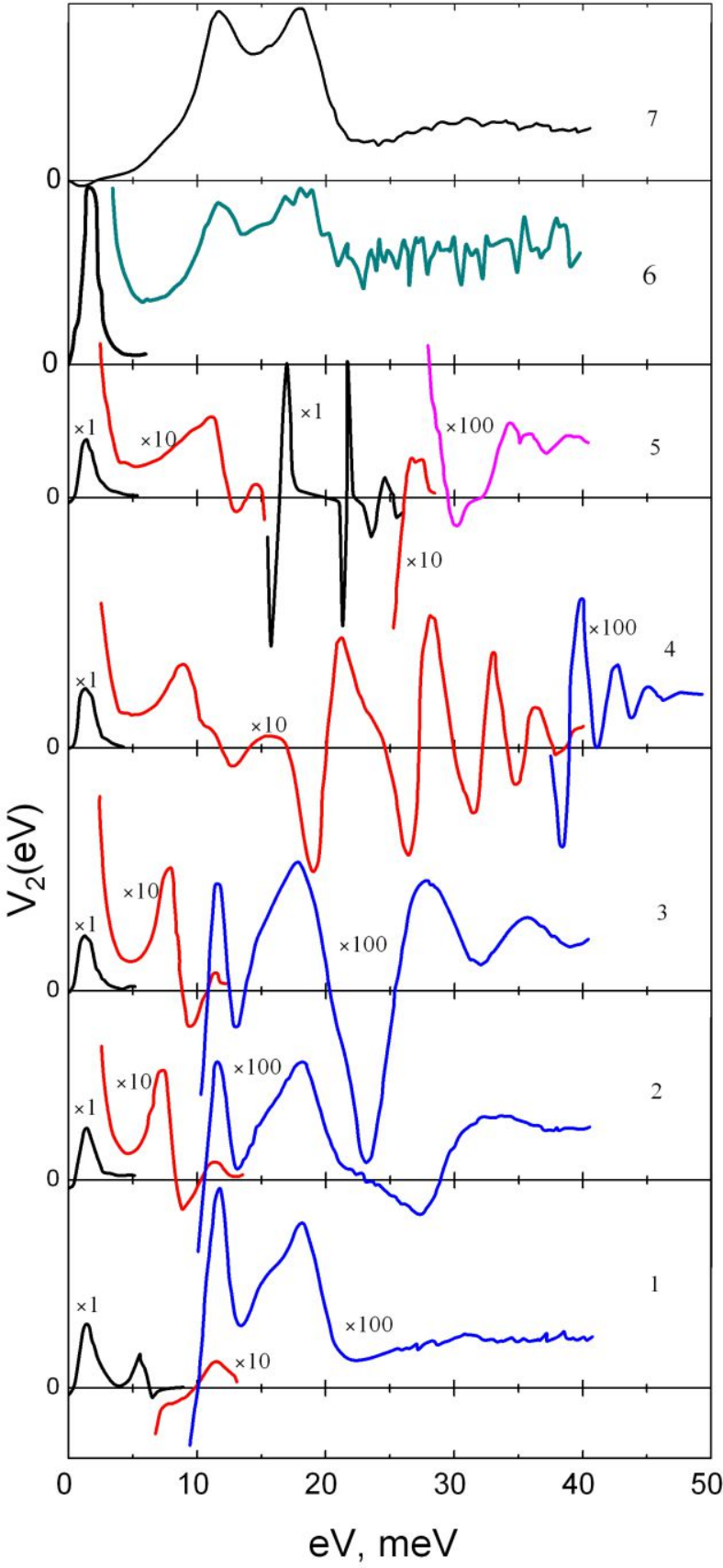}
\caption[] {Field evolution of point contact spectra of $Ta$ with H, Oe: 0(1); 26(2); 32.5 (3); 45.5 (4); 52 (5); 910 (6); 3250 (7). The curve 6 is for a Ta-Cu contact with a remittance of $73 \Omega .$}
\label{Fig2}
\end{figure}
\begin{figure}[]
\includegraphics[width=0.5\textwidth,angle=0]{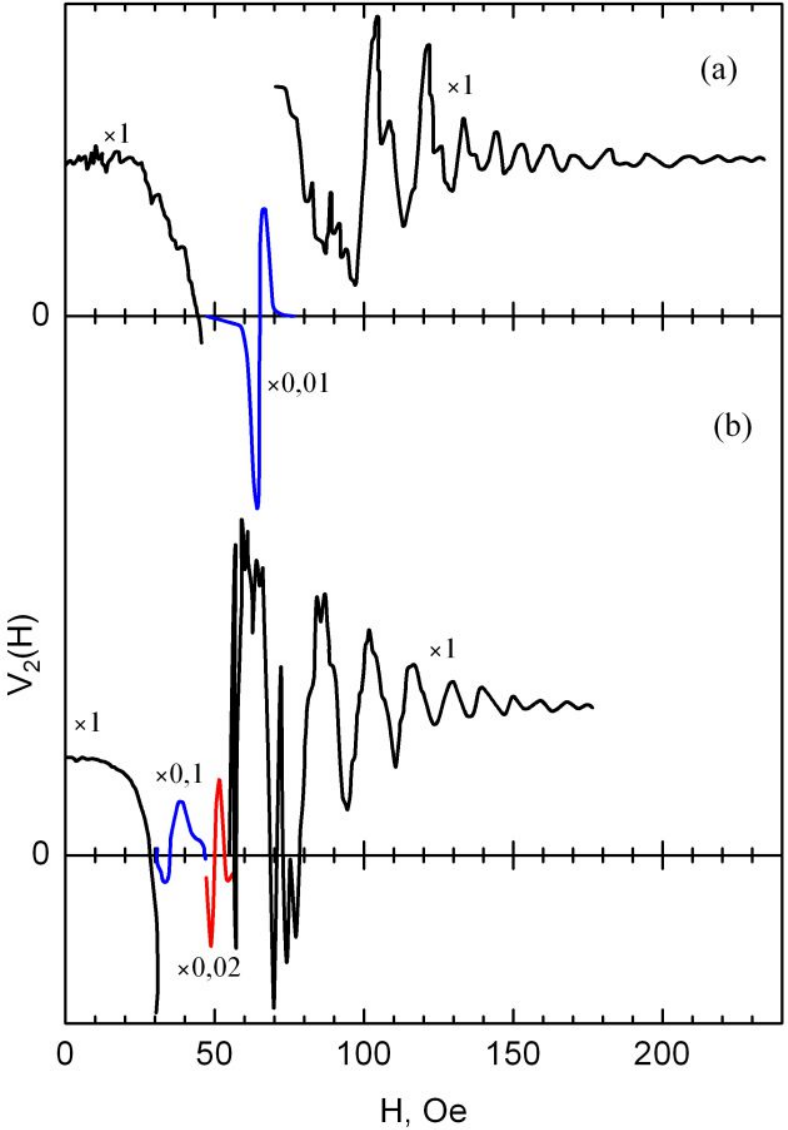}
\caption[] {The curvature of I-V characteristics as a function of magnetic field ($Ta-Cu$ contacts, H being perpendicular to a contact axis): ${{R}_{0}}=60\ \Omega $, $V=11.5\ mV$, $T=1.6\ K(a)$; ${{R}_{0}}=75\ \Omega $, $V=25\ mV$, $T=1.68\ K(b)$.}
\label{Fig3}
\end{figure}
In studying the nonlinearity of IVC of clean $Ta-Cu$ and $Ta-Ta$ point contacts\footnote {Both types of contacts exhibited qualitatively similar features. We present below the results obtained for $Ta-Cu$ contacts.} with dimensions of the order of several tens of angstroms, we discovered that the dependence of the second derivative of the IVC on the bias voltage (the so-called point contact spectrum) changes substantially when $Ta$ transforms into the superconducting state. New features determined by the jump-like change in the properties of the superconductor in the nonequilibrium state (phase transition) at the critical level of power injection appear in the spectrum. The peaks at the phonon energies grow rapidly, and in magnetic fields less than $\sim{\ }0.1\ {{\text{H}}_{\text{cm}}}$  there appear oscillations of the second derivative of the IVC as a function of eV at H=const and as a function of H at eV=const.\\

The contacts were created by the shear method by touching electrodes of dimensions $1.5\times 1.5\times 10~m{{m}^{3}}$ between the side edges. The quality of the point contacts was monitored according to their spectra in the normal state (Figs. \ref{Fig1} and \ref{Fig2}, curves 1 and 7, respectively). The intensity of the spectra reached a value corresponding to ${{\lambda }_{pc}}=0.15$ close to ${{\lambda }_{tr}}=0.13$ for Ta\footnote {The following values of the parameters of Та were used to make the quantitative estimates of the intensity of the spectra and the diameter of the contacts $Ta$: $\rho l=0.96\cdot {{10}^{-12}}\Omega \cdot c{{m}^{2}}$ and ${{v}_{F}}=0.24\cdot {{10}^{8}}cm/\sec $.}. In accordance with the inequality ${{v}_{F}}(\text{Ta})\ll {{v}_{F}}(\text{Cu})$ the EPI in $Cu$ made virtually no contribution to the spectra of the heterocontacts.\\

The evolution of $Ta$ spectra in the $Ta-Cu$ contact with ${{R}_{0}}=64\ \Omega $ and H=0 is shown in Fig. \ref{Fig1}. ${{V}_{2}}(eV)$ is proportional to ${{d}^{2}}V/d{{I}^{2}}(eV)$. All curves are presented on the same scale along the ordinate axis (taking into account the factors changing the scale of different fragments), corresponding to an increase in dV/dI in the normal state by 7.33\% with an increase in the bias from 0 to 23 mV.\\

In a narrow temperature range $\left( {{T}_{c}}-T \right)<0.1\ K$ (not shown in Fig.\ref{Fig1} the intensity of the phonon peaks is much lower, and is rapidly restored as the temperature is further lowered. Then, in the region containing the low-frequency mode of the EPI spectrum in Ta ($\approx 7\ meV$), there arises the feature marked by the arrow in curve 2. The intensity of this feature increases rapidly, and, with a further drop in the temperature, shifts into the region of lower values of eV. For contacts with a high resistance, this feature has a tendency to appear near the $Ta$ peak ($\approx 11.5\ meV$ or the LA peak ($\approx 18\ meV$). Similar features were observed in tin contacts in Ref \cite{Khotkevlch1}, but there they were explained by thermal effects. For $T\lesssim \ 0.5\ {{T}_{c}}$ this feature corresponds to a small (of the order of several percent) decrease in the current and break in the IVC in the direction of the abscissa axis. The phenomena described occur only for contacts with a quite high magnitude of the excess current $({{I}_{exc}}\gtrsim 0.6{{{\Delta }_{0}}}/{{{R}_{0}}}\;$ for $ScN$ and $1.0{{{\Delta }_{0}}}/{{{R}_{0}}}\;$ for $ScS$ contacts). They apparently reflect the behavior of the energy gap, averaged over the region of the nonequilibrium superconductor with dimensions $\sim \xi $, adjoining the point contact. The critical power ${{P}_{c}}={V_{c}^{2}}/{{{R}_{0}}}\;$ at which the phase transition occurs remains approximately constant when the resistance $R_0$ changes by more than anorder of magnitude. It is possible to obtain an estimate for the average critical concentration of excess quasiparticles: ${{n}_{c}}\sim {{J}_{c}}\tau $, where ${{J}_{c}}\sim{\ }{{P}_{c}}/2\Delta {{\xi }^{3}}$ is the rate of generation of quasiparticles per unit volume, while $\tau \sim {\xi }/{{{v}_{F}}}\;$ is the time for them to leave the region of nonequilibrium. Assuming that at T= 2K the magnitude of the gap $\Delta =0.67\cdot {{10}^{-3}}eV$ and $\xi =7\cdot {{10}^{-6}}cm$, we obtain $\sim {{10}^{17}}\div {{10}^{18}}\ c{{m}^{-3}}$ (the typical density at which a jump-like change in the properties of the nonequilibrium state occurs \cite{Dynes1}. In accordance with this model for Sn the critical power must be a factor of ${{\xi }_{Sn}}v_{F,Sn}^{2}/{{\xi }_{Ta}}v_{F,Ta}^{2}\ $ higher, which corresponds to the experiment of Ref \cite{Khotkevlch1}. Since $\xi =\xi \left( l \right)$, where $l=l\left( eV \right)$, the appearance of a new nonequilibrium phase is usually "tied" to regions of more rapid decrease of  $l\left( eV \right)$, i.e., to the maxima of the EPI function.\\

We find additional evidence for the fact that the above-noted feature corresponds to a transition into a new superconducting state by studying the dependence of the point-contact spectra of Та (for а $Ta-Cu$ contact at ${{R}_{o}}=87\ \Omega $ and T=1.6 K) on the magnetic field (Fig. \ref{Fig2}), whose orientation does not qualitatively affect the results.\\

The scale along the ordinate axis corresponds to an increase in   in the normal state by 5.16\% in the interval $0<V<30\ mV$.\\

In fields of the order of only 1 Oe, the critical feature becomes broader, and its intensity decreases substantially. In fields of the order of 10 Oe oscillations of the second derivative of the IVC appear, which "creep" into the EPI spectrum from the side of high values of eV with increasing H (curves 2 and 3). Since the position of the critical feature shifts toward higher values of eV almost linearly with increasing H, while the position of the first oscillation moves opposite to it according to the same law, both types of features soon coalesce masking the phonon peaks (curves 4 and 5). In a magnetic field of $\sim{\ }{{H}_{cm}}$ the traces of superconductivity in the spectrum still remain In the form of relatively weak gap feature near V=0 and fluctuations in the region of large biases (curve 6), and the point-contact spectrum corresponds completely to the normal state (curve 7) only in fields of $2.5\div 3$ kOe. It is interesting to note that the form of the oscillations in ${{V}_{2}}\left( eV \right)$ in the curves 2-5 corresponds to a step-like increase in the current with increasing eV.\\

Figures \ref{Fig3} (a) and (b) show the dependences ${{V}_{2}}(H)$ with two biases eV, corresponding to the conditions $eV\gtrsim \hbar {{\omega }_{D}}$ and $eV\approx \hbar {{\omega }_{TA}}$ respectively.\\

Thus the facts presented above indicate that a new nonequilibrium superconducting state with an average gap of the order of ${{\Delta }_{0}}$ appears near the point contact. The nature of the inelastic electron and phonon scattering processes leading to the nonlinearities in the IVC differ substantially in this state from the corresponding processes in the normal metal. It is possible that in this case phonons with a low group velocity which cannot leave the region of the contact, enable, by being reabsorbed by Cooper pairs, the multiplication of quasiparticles, which is what causes the sharpening of the peaks at the corresponding energies. This circumstance can be used in studying the fine details of the EPI in superconductors, including materials with high critical parameters.\\
\begin{center}
\textbf{NOTATION}
\end{center}

Here $H_{cm}$ is the thermodynamic magnetic field, $\lambda _{pc}$ is the point-contact parameter of the electron-phonon interaction (EPI), $\lambda _{tr}$ is the transport parameter of the EPI, $p_c $ is the critical power, $n_c$ is the critical concentration,  $V_c$ is the critical voltage, and $J_c$ is the generation rate.

\end{document}